# Interface-induced collective phase transition in $VO_2$-based bilayers studied by layer selective spectroscopy

D. Shiga[1,2,*], S. Inoue[1], T. Kanda[1], N. Hasegawa[1], M. Kitamura[2], K. Horiba[2], K. Yoshimatsu[1], A. F. Santander-Syro[3], and H. Kumigashira[1,2,*]

[1] *Institute of Multidisciplinary Research for Advanced Materials (IMRAM), Tohoku University, Sendai 980–8577, Japan*

[2] *Photon Factory, Institute of Materials Structure Science, High Energy Accelerator Research Organization (KEK), Tsukuba 305–0801, Japan*

[3] *Institut des Sciences Moléculaires d'Orsay, Université Paris-Saclay, 91405 Orsay, France*

**Abstract**

We investigated the origin of collective electronic phase transitions induced at the heterointerface between monoclinic insulating $VO_2$ and rutile metallic electron-doped $VO_2$ layers using *in situ* soft x-ray photoemission spectroscopy (PES) and x-ray absorption spectroscopy (XAS) on nanoscale $VO_2/V_{0.99}W_{0.01}O_2$ $(001)_R$ bilayers. Thanks to the surface sensitivity of PES and XAS, we determined the changes in the electronic structure and V-V dimerization in each constituent layer separately. The layer selective observation of the electronic and crystal structures in the upper $VO_2$ layer of the bilayer indicates that the monoclinic insulating phase $VO_2$ layer undergoes a transition to the rutile metallic phase by forming the heterointerface. Detailed temperature-dependent measurements reveal that the rutile metallic phase $VO_2$ undergoes a transition to the monoclinic insulating phase with a decrease in temperature, as in the case of a $VO_2$ single-layer film. Furthermore, during the temperature-induced phase transition in the $VO_2$ layer, the spectra are well described by an in-plane phase separation of the rutile metallic and monoclinic insulating phases. These results suggest that the interface-induced transition from the monoclinic insulating to the rutile metallic phase in the $VO_2$ layer of bilayers occurs as a collective phase transition derived from the static energy balance between the interfacial energy and the bulk free energies of the constituent layers.

*Correspondence authors: dshiga@tohoku.ac.jp, kumigashira@tohoku.ac.jp

## I. INTRODUCTION

Vanadium dioxide ($VO_2$) shows an intriguing first-order metal-insulator transition (MIT) near room temperature that has been controversially discussed for over 60 years, particularly because both a structural transition and electron correlations contribute to it [1–17]. Across the MIT, the crystal and electronic structures change from a high-temperature rutile metallic phase to a low-temperature monoclinic insulating phase, concomitant with a dimerization of V atoms along the $[001]_R$ ($c_R$-axis) direction [2,3]. The MIT is also accompanied by a change in conductivity of several orders of magnitude and exhibits an ultrafast response to external stimuli. Thus, the MIT in $VO_2$ has become a central topic in condensed matter physics for its potential applications in Mottronics devices [18–24]. Further, this unusual phenomenon originating from the interplay of lattice dynamics and electron correlations provides a unique testbed to understand the physics of strongly correlated oxides.

The structural phase transition in $VO_2$ is characterized by the tilting and pairing of V ions, resulting in the collective dimerization of V ions along the $[001]_R$ direction in the monoclinic insulating phase [2,3]. Although the MIT that is concomitant with the collective V-V dimerization is reminiscent of the Peierls transition [4,5], the important role that strong electron correlations have in driving the MIT in $VO_2$ has also been evidenced from a large number of experimental and theoretical investigations [6,7]. Therefore, the mechanism of the MIT in $VO_2$ is now mainly understood as a cooperative Mott-Peierls transition [9–17].

This type of MIT in $VO_2$ has motivated researchers to control the electronic and structural phase transition by changing the delicate interplay between the electron correlations and the lattice via interfacial effects. Yajima *et al*. [25] have reported the occurrence of a collective MIT in bilayer structures composed of $VO_2$ layers with different MIT temperatures ($T_{\mathrm{MIT}}$s). The bilayer consists of an undoped $VO_2$ layer with $T_{\mathrm{MIT}}$ = 293 K and an electron-doped (W-doped) $VO_2$ layer with a slightly lower $T_{\mathrm{MIT}}$ of 257 K. It shows a collective (interlocked) transition at a certain critical layer thickness, while each layer undergoes the MIT at its respective $T_{\mathrm{MIT}}$ above the collective length, which corresponds to ~4.5 nm for each layer. In the collective MIT, the $VO_2$ layer undergoes a transition from insulator to metal by forming the interface with the electron-doped $VO_2$. This collective transition is interpreted as originating from the static energy balance between the interfacial energy and the bulk free energies (the electronic and structural energies) of the constituent layers based on detailed thermodynamic calculations [25]. In this scenario (Scenario I), the collective phase transition of the $VO_2$ layer is interpreted as increased stability of the rutile metallic phase relative to the monoclinic insulating phase to avoid the cost of interfacial energy between the different phases.

In the collective MIT derived from the static energy balance, the MIT in the $VO_2$ layer should be the original transition from the monoclinic insulating to the rutile metallic phase. Namely, the metallic state in the $VO_2$ layer should be in the rutile metallic phase. However, Lee *et al.* [26] have recently suggested the possibility of realizing a new "monoclinic metallic" phase in the $VO_2$ layer of the bilayer structure. In this alternative scenario (Scenario II), an isostructural transition from monoclinic insulating to monoclinic metallic phase, namely, a pure electronic phase transition without any structural transition, occurs in the $VO_2$ layer adjacent to the electron-doped $VO_2$ layer, although the doped layer maintains the rutile metallic phase.

The interface-induced collective phase transition occurring in $VO_2$-based bilayers remains an active subject of both experimental and theoretical research [25–28]. However, the lack of conclusive experimental proof regarding which metallic phase actually emerges has hindered the understanding of the phase transition, particularly in theoretical investigations of how interfacial effects can stabilize novel electronic phases or influence the Mott-Peierls transition pathway [27,28]. Therefore, there is a critical need to experimentally clarify the electronic and crystal structures of the $VO_2$ layer in the bilayer, while eliminating the influence from the counterpart layer [26].

To verify which of the two metallic states [rutile metallic phase (Scenario I) or monoclinic metallic phase (Scenario II)] is realized in the $VO_2$ layer, it is crucial to observe the electronic and crystal structures in each constituent layer separately. Against this background, in this study, we employed surface-sensitive photoemission spectroscopy (PES) and x-ray absorption spectroscopy (XAS) measurements in the soft x-ray region to investigate the change in the electronic and crystal (characteristic V-V dimer) structures in the top $VO_2$ layer of $VO_2/V_{0.99}W_{0.01}O_2$ (W:$VO_2$) $(001)_R$ bilayer structures. Thanks to a sufficiently shallow probing depth (1.5–2 nm) compared to each layer thickness (4.5 nm) in both spectroscopic measurements [15], information on the electronic and crystal structures of the upper $VO_2$ layer was selectively extracted. The PES and XAS spectra exhibited remarkable changes associated with phase transition in the upper $VO_2$ layers: (1) The upper $VO_2$ layer exhibits almost the same spectral behavior across the MIT as that of a $VO_2$ single-layer film, whereas its $T_{MIT}$ is slightly lower than that of the single-layer film. (2) In the metallic states of the upper $VO_2$ layer, there is no indication of the V-V dimerization. (3) During the temperature-induced phase transition, both the PES and XAS spectra are described by a linear combination of the rutile metallic and monoclinic insulating phases, indicating the occurrence of an in-plane phase separation. These results strongly suggest that the upper $VO_2$ layer undergoes a collective transition from the monoclinic insulating to the rutile metallic phase by forming the heterointerface with the electron-doped $VO_2$ layer and that the monoclinic metallic phase does not emerge in the present $VO_2$/W:$VO_2$ $(001)_R$ heterostructures. The occurrence of the phase transition from the monoclinic insulating to the rutile metallic phase in the $VO_2$ upper layers suggests that the collective phase transition originates from the static energy balance between the interfacial energy and the bulk free energies of the constituent layers (Scenario I).

## II. EXPERIMENT

The $VO_2/V_{0.99}W_{0.01}O_2$ heterostructures were fabricated on the (001) surface of 0.05 wt% Nb-doped rutile-$TiO_2$ substrates in a pulsed-laser deposition (PLD) chamber connected to an *in situ* PES system at BL-2A MUSASHI of the Photon Factory, KEK [15–17,29,30]. Sintered pellets with appropriate compositions of $V_2O_5$ and $V_{1.98}W_{0.02}O_5$ were used as PLD ablation targets. Each layer was grown at a rate of 0.02 nm $s^{-1}$, as estimated from the Laue fringes in x-ray diffraction (XRD) patterns of a corresponding single-layer film. The growth conditions for each layer are described in detail elsewhere, and the characterization of the heterostructures is given in Supplemental Material [31]. During the deposition, the substrate temperature was maintained at 400 °C, and the oxygen pressure was maintained at 10 mTorr. The thicknesses of the deposited $VO_2$ and W:$VO_2$ layers, as well as those of the $VO_2$ and W:$VO_2$ single-layer films, were precisely controlled by deposition time. The surface structures and cleanliness of the heterostructures were confirmed via reflection high-energy electron diffraction (see Fig. S1 in Supplemental Material [31]) and core-level photoemission measurements (see Fig. S3 in Supplemental Material [31]), respectively. The formation of a chemically abrupt interface in the present bilayers was also confirmed by detailed analysis of core-level intensities based on the photoelectron attenuation function (see Fig. S3 in Supplemental Material [31]).

The surface morphologies of the measured heterostructures and single-layer films were analyzed via atomic force microscopy in air (see Fig. S1 in Supplemental Material [31]). The epitaxial relationship and crystalline quality were characterized by XRD, confirming the coherent growth of each layer. A sharp diffraction pattern with well-defined Laue fringes was observed, indicating the high quality of the heterostructures, i.e., homogeneously coherent films with atomically flat surfaces and interfaces (see Fig. S2 in Supplemental Material [31]). All bilayer structures were fabricated under the same conditions as previously reported bilayer structures [25], wherein chemically abrupt interfaces formed. The sheet conductance was measured using the standard four-probe method with a temperature ramp rate of 10 mK $s^{-1}$.

PES measurements were performed *in situ* with the use of a VG-Scienta SES-2002 analyzer with a total energy resolution of 120 meV at a photon energy of 700 eV. The vacuum transferring of the grown samples was necessary to prevent the overoxidation of the surface layer [15–17]. The XAS spectra were also measured *in situ* with linearly polarized light via the measurement of the sample drain current. For linear dichroism measurement of oxygen *K*-edge XAS (O *K* XAS), we acquired the XAS spectra at angles $\theta = 0°$ and 60° between the $[100]_R$ direction and the polarization vector $\boldsymbol{E}$ while maintaining a fixed angle between the direction normal to the interfaces and the incident light (see Fig. S4 in Supplemental Material [31]). Here, we emphasize that the present O *K* XAS measurement is also surface sensitive; its probing depth is estimated to be almost identical to that of the present soft x-ray PES measurements (1.5–2 nm) (see Fig. S5 in Supplemental Material [31]). Therefore, both the PES and XAS spectra reflect the information from almost the same surface region of the top $VO_2$ layers. The Fermi energy ($E_F$) of each sample was determined by the measurement of a gold film that was electrically connected to the sample. As it is common knowledge that $VO_2$ exhibits an insulator-to-metal transition upon irradiation by light [35], we paid particular attention to the possible spectral changes induced by light irradiation (see Fig. S6 in Supplemental Material [31]). The stoichiometry of the constituent layers was carefully characterized by analyzing the relative intensities of the V-2*p* and -3*p*, O-1*s*, and W-4*f* core levels, confirming that the cation composition of the samples was the same as that of the PLD ablation targets. We carefully carried out temperature-dependent PES and XAS measurements by confirming that the spectral changes with temperature were saturated (hysteresis effects [11,25,32,39] were no longer present). Furthermore, the sample temperature was maintained at 320 K for half an hour before measuring, and then the measurements were performed only upon cooling to avoid the possible hysteresis effect (see Fig. S7 in Supplemental Material [31]).

## III. RESULTS & DISCUSSION

Before discussing the PES spectra, we provide evidence that the prepared bilayers show essentially the same properties as those in the previous report [25]. Figure 1 shows the temperature dependence of sheet conductivity $\sigma_{\mathrm{Sheet}}$ upon cooling for $VO_2$ (4.5 nm)/W:$VO_2$ (4.5 nm) and $VO_2$ (6.5 nm)/W:$VO_2$ (6.5 nm) bilayers, together with that for 9 nm $VO_2$ and W:$VO_2$ single-layer films. For the single-layer films, the $\sigma_{\mathrm{Sheet}}$-$T$ curve steeply changes across the MIT [15–17,43,44], as reported previously [25]. The corresponding $T_{\mathrm{MIT}}$ on the cooling process ($T_{\mathrm{MIT}}^{\mathrm{Cooling}}$), which is defined as the inflection point of each $\log_{10}\sigma_{\mathrm{Sheet}}$-$T$ curve, is determined to be 286 K and 261 K for the $VO_2$ and W:$VO_2$ single-layer films, respectively. The values are in excellent agreement with the previous reports on epitaxial films coherently grown on $TiO_2$ (001) substrates, guaranteeing almost the same qualities of the present films as those of the previous ones [15–17,23,25,32,39]. For the bilayers, two separate transitions are observed, reflecting the original $T_{\mathrm{MIT}}^{\mathrm{Cooling}}$ of constituent layers. This result suggests that the constituent layers behave independently despite the layered structure, although the $T_{\mathrm{MIT}}^{\mathrm{Cooling}}$ corresponding to the $VO_2$ layer is slightly lower. In contrast, the two transitions seem to merge at 4.5 nm bilayers, reflecting the occurrence of the collective transition throughout the whole bilayer structure in the temperature range of 260–275 K. The observed layer-thickness dependence is in good agreement with that in the previous study [25], indicating the achievement of essentially the same properties in the present bilayers. Furthermore, this good agreement indicates the negligible influence of W interdiffusion on the observed phenomena in the present bilayers, since the formation of a chemically abrupt interface has been confirmed by the previous study [25]. Note that our detailed analysis for W 4*f* core-level spectra also supported the formation of a chemically abrupt interface (i.e., the absence of detectable interdiffusion of W ions) in the present study (Fig. S3 in Supplemental Material [31]). It should also be noted that the crystallinity of the measured single-layer films and bilayers is almost identical to that in the previous study [25] (see Figs. S1 and S2 [31]).

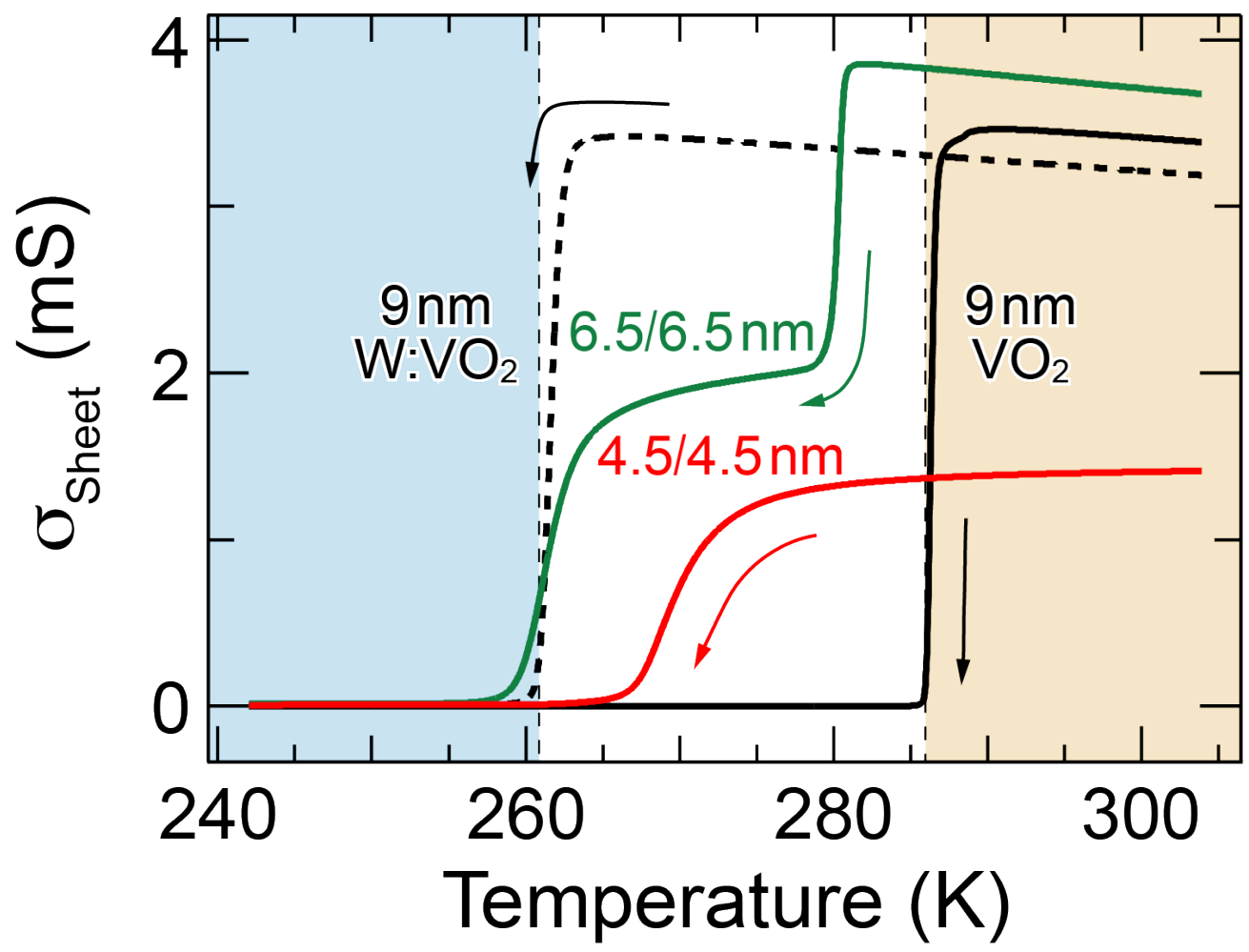

**FIG. 1.** Temperature dependence of $\sigma_{\mathrm{Sheet}}$ upon cooling for $VO_2$ (4.5 nm)/W:$VO_2$ (4.5 nm) bilayer (red curve) and $VO_2$ (6.5 nm)/W:$VO_2$ (6.5 nm) bilayer (green curve), together with that of 9 nm $VO_2$ and W:$VO_2$ single-layer films (solid and dashed black curves, respectively). Vertical dashed lines indicate $T_{\mathrm{MIT}}^{\mathrm{Cooling}}$ of the $VO_2$ and electron-doped W:$VO_2$ single-layer films (286 and 261 K, respectively). Here, $T_{\mathrm{MIT}}^{\mathrm{Cooling}}$ is defined as the inflection point in the $\log_{10}\sigma_{\mathrm{Sheet}}$-$T$ curve in the cooling process (see text in more detail). The 6.5 nm bilayer exhibits two-separate transitions, indicating that the upper and lower layers undergo MIT separately. In contrast, the 4.5 nm bilayer shows a single transition, indicating the occurrence of the collective transition in the temperature range of 260–275 K. Note that the behaviors are essentially the same as the previous reports, guaranteeing the comparable quality of the single-layer films and bilayers in the present study [25]. Color shading is a guide for the eye to visually separate specific temperature regions corresponding to the expected electronic phases in the 4.5 nm bilayer, as schematically illustrated in the upper panel of Fig. 2, where both layers are in either the rutile metallic (light orange) or monoclinic insulating (light blue) phase (see text for more details).

To see the behavior of the collective phase transition in more detail, we present the temperature dependence of the sheet resistance $R_{\mathrm{Sheet}}$ in Fig. 2, alongside reference data of $VO_2$ and W:$VO_2$ single-layer films. The 4.5 nm bilayer exhibits a single-step MIT at $T_{\mathrm{MIT}}^{\mathrm{Cooling}} \sim 267$ K, in sharp contrast to the two-step MIT in the 6.5 nm bilayer (see Fig. 1). Furthermore, the transition is relatively broad in comparison with the abrupt change across the MIT in the single-layer films, implying a complicated interplay between different phases in the two layers.

According to the previous studies [25,26], the complicated behaviors may be classified into four temperature regions *A*–*D*, as schematically illustrated in the upper panel of Fig. 2. At temperature *A*, both layers are in the rutile metallic phase, while at temperature *D* both layers are in the monoclinic insulating phase, since at those temperatures the original $VO_2$ and W:$VO_2$ layers are in the same corresponding rutile metallic or monoclinic insulating phases. Meanwhile, the upper $VO_2$ layer shows metallic behavior at temperature *B* as a result of the collective phase transition, although the $VO_2$ (W:$VO_2$) layer should be the monoclinic insulating (rutile metallic) phase in the case that each of the layers behaves independently as in the 6.5 nm bilayer. In other words, the $VO_2$ layer exhibits an insulator-to-metal transition by forming the interface with the electron-doped $VO_2$ with lower $T_{MIT}$. For Scenarios I and II, the possible metallic phases that emerged at *B* due to the collective interface-induced phase transition are rutile metal and monoclinic metal, respectively. With decreasing the temperature, the metallic phase eventually exhibits a temperature-induced MIT around temperature *C*. As a result, the $T_{MIT}$ of the $VO_2$ (W:$VO_2$) layer is suppressed (enhanced) owing to the proximity effect between the two layers.

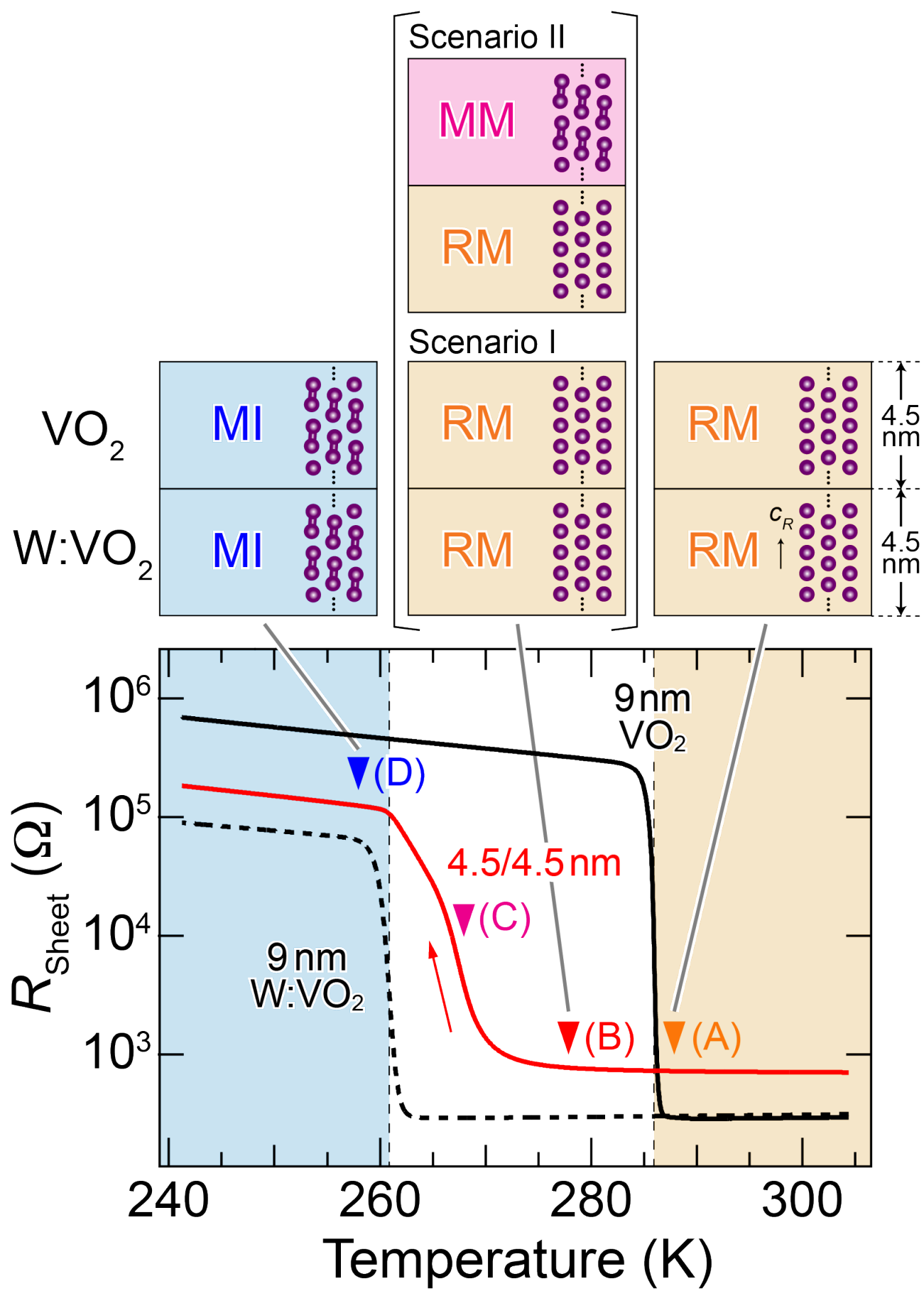

**FIG. 2.** Temperature dependence of $R_{\mathrm{Sheet}}$ measured upon cooling for $VO_2$ (4.5 nm)/W:$VO_2$ (4.5 nm) bilayer (solid red curve), along with those of 9 nm $VO_2$ and W:$VO_2$ single-layer films (solid and dashed black curves, respectively). Vertical dashed lines indicate $T_{\mathrm{MIT}}^{\mathrm{Cooling}}$ of the $VO_2$ and W:$VO_2$ single-layer films (286 and 261 K, respectively), while colored solid triangles indicate the spectroscopic measurement points (*A*–*D*). The upper panel shows the schematic illustration for the expected electronic phases of the bilayer at temperatures *A*, *B*, and *D*. At temperature *B*, the two possible metallic phases of the upper $VO_2$ layer based on Scenarios I (interface-induced collective phase transition from monoclinic insulator to rutile metal [25]) and II (isostructural interface-induced collective phase transition from monoclinic insulator to monoclinic metal [26]) are presented. Color shading is a guide for the eye to visually separate specific temperature regions corresponding to the expected electronic phases in the 4.5 nm bilayer, as schematically illustrated in the upper panel, where both layers are in either the rutile metallic (light orange) or monoclinic insulating (light blue) phase (see text for more details). MI, RM, and MM denote the monoclinic insulator, rutile metal, and monoclinic metal, respectively.

To identify the metallic phase at temperature *B*, we measured the changes in the electronic structures and V-V dimerization in the upper $VO_2$ layer selectively through PES measurements. Figure 3(a) shows the valence-band spectra measured upon cooling at temperatures *A*–*D* for the $VO_2$ (4.5 nm)/W:$VO_2$ (4.5 nm) bilayers grown on Nb:$TiO_2$ (001) substrates, in addition to those of $VO_2$ single-layer films in the monoclinic insulating and rutile metallic phases as references. Owing to the surface sensitivity of soft x-ray PES, the spectra mostly reflect the electronic structure of the top 4.5 nm $VO_2$ layers, as schematically illustrated in the inset, since the probing depth of the present PES is 1.5–2 nm [15,31]. The spectra contain two main features: structures derived from O 2*p* states at binding energies of 3–10 eV, and peaks derived from the V 3*d* states near $E_F$. The spectra of the bilayer (i.e., the top $VO_2$ layer) exhibit the characteristic features of the MIT in $VO_2$ films [11,15–17,41–43]: the spectrum near $E_F$ at temperature *A* consists of a sharp coherent peak at $E_F$ and a weak broad satellite around 1.2 eV. Meanwhile, the spectrum at temperature *D* shows a single peak around 0.8 eV, corresponding to the formation of an energy gap at $E_F$.

For the upper $VO_2$ of the bilayer structure, the valence-band spectra at temperatures *A* and *D* are almost exactly the same as those of the $VO_2$ single-layer films in the rutile metallic and monoclinic insulating phases, respectively. At temperature *B* where the $VO_2$ layer changes from the original monoclinic insulating phase to the metallic phase (rutile metal in Scenario I or monoclinic metal in Scenario II) by forming the heterointerface (see Fig. 2), the spectra exhibit the features representative of the rutile metallic phase of the $VO_2$ single-layer film. In fact, the spectrum at temperature *B* is identical to the spectrum at temperature *A* (rutile metallic phase) within the experimental error, suggesting the occurrence of the monoclinic insulator-to-rutile metal transition by forming the interface with the W:$VO_2$ layer. The metallic state eventually exhibits the transition to the monoclinic insulating phase at temperature *D* through certain complicated phenomena around temperature *C* (see Fig. 2). Furthermore, focusing on the O 2*p* states, we observe dramatic changes across the temperature-dependent MIT. For $VO_2$ films, these changes are attributed to the structural changes (V-V dimerization) concomitant with the MIT in $VO_2$ [11,15–17], strongly suggesting the occurrence of temperature-induced structural transition from the rutile metallic phase (temperatures *A* and *B*) to the monoclinic insulating phase (temperature D) in the upper $VO_2$ layer.

The next crucial issue is whether the crystal structure of the interface-induced metallic phase (see

Supplemental Note 7 [31]) is the original monoclinic structure with the V-V dimerization or the rutile structure without the dimerization. To determine the crystal structure (monoclinic or rutile) of the interface-induced metallic phase, as well as those of the other phases, we measured, as shown in Fig. 3(b), the polarization dependence of O *K* XAS, which has been previously used as an indicator of V-V dimerization. The O *K* XAS probes the unoccupied O 2*p* partial density of states that are mixed with the unoccupied V 3*d* states and is thus complementary to PES for investigating the electronic structures of conduction bands. Because the V-V dimerization in the monoclinic insulating phase splits the half-filled $d_{||}$ state into occupied $d_{||}$ and unoccupied $d_{||}^*$ states, an additional peak corresponding to the $d_{||}^*$ states appears in the XAS spectra only for the monoclinic insulating phase [3,11,30,44–46]. Moreover, owing to strict dipole selection rules, the additional $d_{||}^*$ states only appear in the spectra acquired with $\boldsymbol{E}$ parallel to the $[001]_R$ ($c_R$-axis) direction ($\boldsymbol{E} \parallel c_R$). From the assignments made in previous studies [11,44], the $d_{||}^*$ peak emerges at 530.4 eV in the monoclinic insulating phase (temperature *D*) measured with the $\boldsymbol{E} \parallel c_R$ geometry, whereas it disappears in the rutile metallic one (temperature *A*). The identification of the $d_{||}^*$ states was further confirmed by the linear dichroism of the XAS spectra: the additional $d_{||}^*$ peak in the monoclinic insulating phase disappeared for the spectrum taken with $\boldsymbol{E} \perp c_R$ (see Fig. S4 in Supplemental Material [31]). Thus, the existence of the $d_{||}^*$ peak in the spectra with the $\boldsymbol{E} \parallel c_R$ geometry can be used as a fingerprint of the V-V dimerization in monoclinic $VO_2$ [3,11,44,45]. This well-established determination method based on the presence or absence of the $d_{||}^*$ peak provides a reliable indicator to determine whether the interface-induced phase retains the dimerized monoclinic metallic phase or becomes the nondimerized rutile metallic phase.

Figure 3(b) shows the temperature dependence of *in situ* O *K* XAS spectra upon cooling acquired with $\boldsymbol{E} \parallel c_R$ of the bilayers. It should be noted that since the probing depth of the present XAS measurements is comparable to that of the PES measurements (Fig. S5 in Supplemental Material [31]), both spectra reflect information on the top 4.5 nm $VO_2$ layers [see the inset of Fig. 3(a)]. The spectra for the upper $VO_2$ layer show the rutile metallic and monoclinic insulating nature at temperatures *A* and *D*, respectively, as in the case of $VO_2$ single-layer films. In addition, the spectral shapes at the two temperatures *A* (rutile metal) and *B* are almost identical, and no detectable $d_{||}^*$ state is observed, indicating the absence of V-V dimerization at temperature *B*. The selective observation of the electronic and crystal structures indicates that the upper $VO_2$ layer undergoes the transition from the monoclinic insulating to the rutile metallic phase by forming the heterointerface (Scenario I).

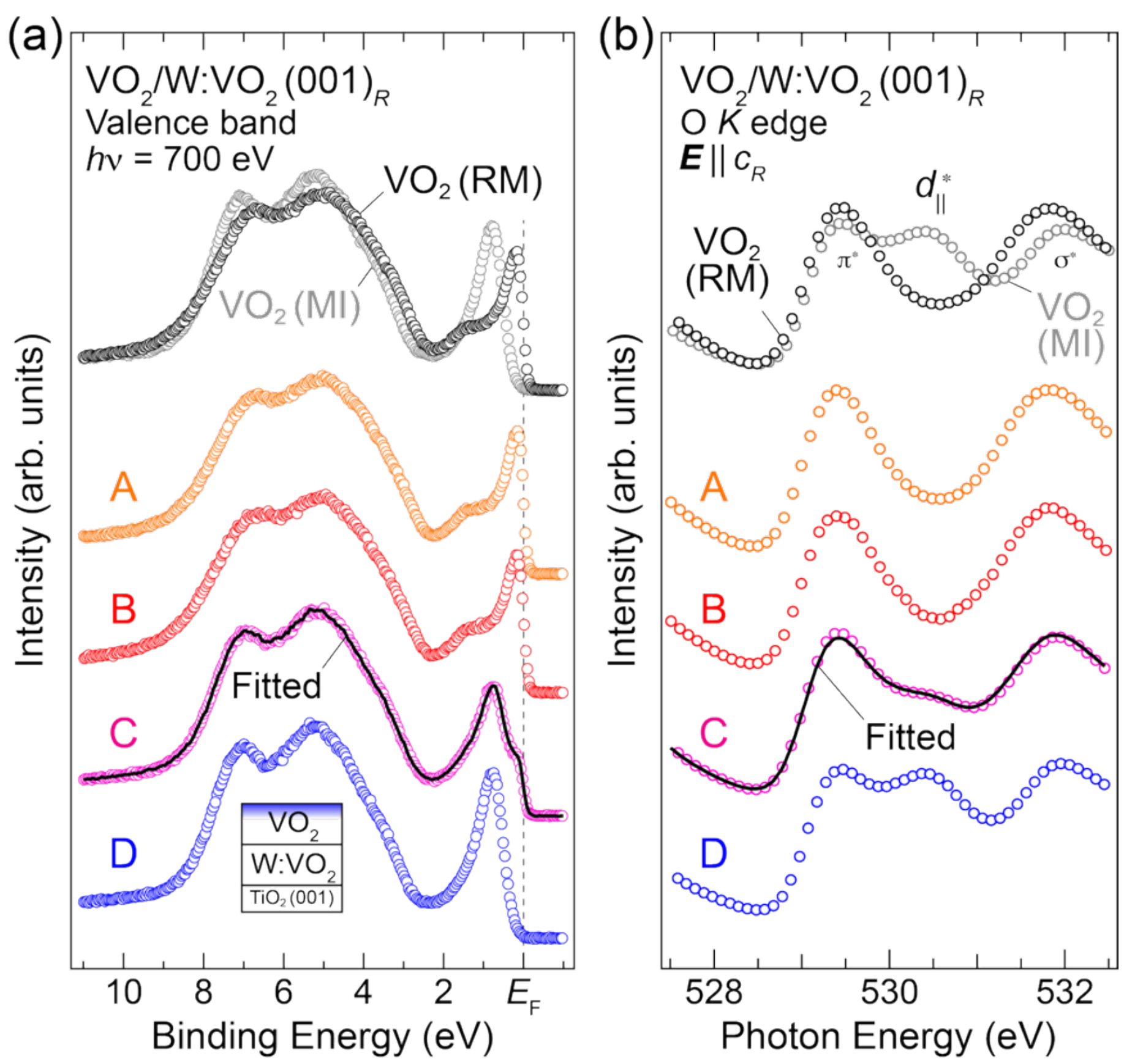

**FIG. 3.** Temperature dependence of *in situ* (a) valence-band spectra and (b) O *K* XAS spectra acquired with the ***E*** ∥ $c_R$ geometry [31] measured upon cooling for $VO_2$ (4.5 nm)/$W:VO_2$ (4.5 nm) bilayers, together with a $VO_2$ film ($T_{MIT}^{Cooling}$ = 286 K) measured at its rutile metallic and monoclinic insulating phases as references. The measured temperatures of *A*, *B*, *C*, and *D* are shown in Fig. 2. For spectra *C*, the fitted result by the linear combination of the rutile metallic (*A*) and monoclinic insulating (*D*) phases [Eq. (1)] is overlaid by a black solid curve. Note that since the probing depth of the present (a) PES and (b) XAS measurements are comparable (1.5–2 nm [15,31]), both spectra reflect information on the top $VO_2$ layers, as schematically illustrated in the inset in (a). The spectra at temperature *B* are almost the same as those of a $VO_2$ film in the rutile metallic phase, as well as those at temperature *A*, indicating that the upper $VO_2$ layer undergoes the transition from monoclinic insulator to rutile metal by forming the heterointerface. MI and RM denote the monoclinic insulator and rutile metal, respectively.

In the $R_{Sheet}$-$T$ curve in Fig. 2, the temperature width of the collective MIT for the 4.5 nm $VO_2/W:VO_2$ bilayer is much broader than that of the original $VO_2$ and $W:VO_2$ single-layer films.

Such broadening was also observed in the previous bilayers [25] as well as the $VO_2$ films grown on $TiO_2$ (110) substrates [35,47,48], implying the occurrence of some in-plane phase separation characteristic of $VO_2$ nanostructures. To shed light on the phenomena, we measured the PES and XAS spectra at temperature *C*, as shown in Fig. 3. The valence-band spectra at temperature *C* near $E_F$ exhibit the peak at 0.8 eV, the same binding energy of spectra *D* corresponding to the monoclinic insulating phase, and an additional structure at $E_F$ indicative of metallic behavior. Upon closer inspection, spectra *C* appear to be an average of the spectra measured at temperatures *A* (rutile metallic phase) and *D* (monoclinic insulating phase). In general, in the case of phase separation, as the size of the soft x-ray light spot used in the present PES and XAS experiments is much larger than the typical size of the phase domains, the measured spectrum is described by a linear combination of the spectra of each phase. Thus, we fit the spectra $I(\alpha)$ using the following equation [17,49]:

$$I(\alpha) = \alpha\, I_{\mathrm{RM}} + (1-\alpha)\, I_{\mathrm{MI}}, \tag{1}$$

where $\alpha$ is the fraction of the rutile metallic phase, and $I_{RM}$ and $I_{MI}$ are the spectra of the rutile metallic phase (spectra *A*) and monoclinic insulating phase (spectra *D*), respectively. As can be seen at temperature *C*, both the PES and XAS spectra are almost perfectly described by the linear combination of the spectra for *A* (rutile metallic phase) and *D* (monoclinic insulating phase) with $\alpha = 0.4$–$0.5$. These results indicate that a phase separation of the rutile metallic and monoclinic insulating phases occurs in the upper $VO_2$ layer of the $VO_2$/W:$VO_2$ bilayer structure.

To investigate the phase-separation behavior in more detail, we measured the detailed temperature dependence of the PES and XAS spectra across the MIT and fitted them using Eq. (1), as shown in Fig. 4. As can be seen in Figs. 4(a) and 4(b), all the spectra during the phase transition are well described by the linear combination. The estimated values of $\alpha$ are plotted as a function of measured temperature in Fig. 4(c), together with the corresponding $\sigma_{\mathrm{Sheet}}$ (Fig. 1). Although the electrical conductance in the case of phase separation must be considered in terms of a percolation model, the good agreement between the $\sigma_{\mathrm{Sheet}}$ and $\alpha$ values suggests that the complex behavior across the MIT in the bilayer can be attributed to the phase separation. Furthermore, considering the difference in probing depth between these two types of measurements (soft x-ray spectroscopies are sensitive to the top $VO_2$ layer, while resistivity measurement probes the entire bilayer), it is likely that the phase domain structures of the upper and lower layers in the bilayers are the same and that the bilayers undergo the collective phase transition. In such a case where the separation of the rutile metallic and monoclinic insulating phases occurs, it may appear that the monoclinic metallic phase emerges. Therefore, it is important to investigate the electronic and V-V dimer structures of each layer selectively using spectroscopic techniques that can prove only specific regions, as in this study.

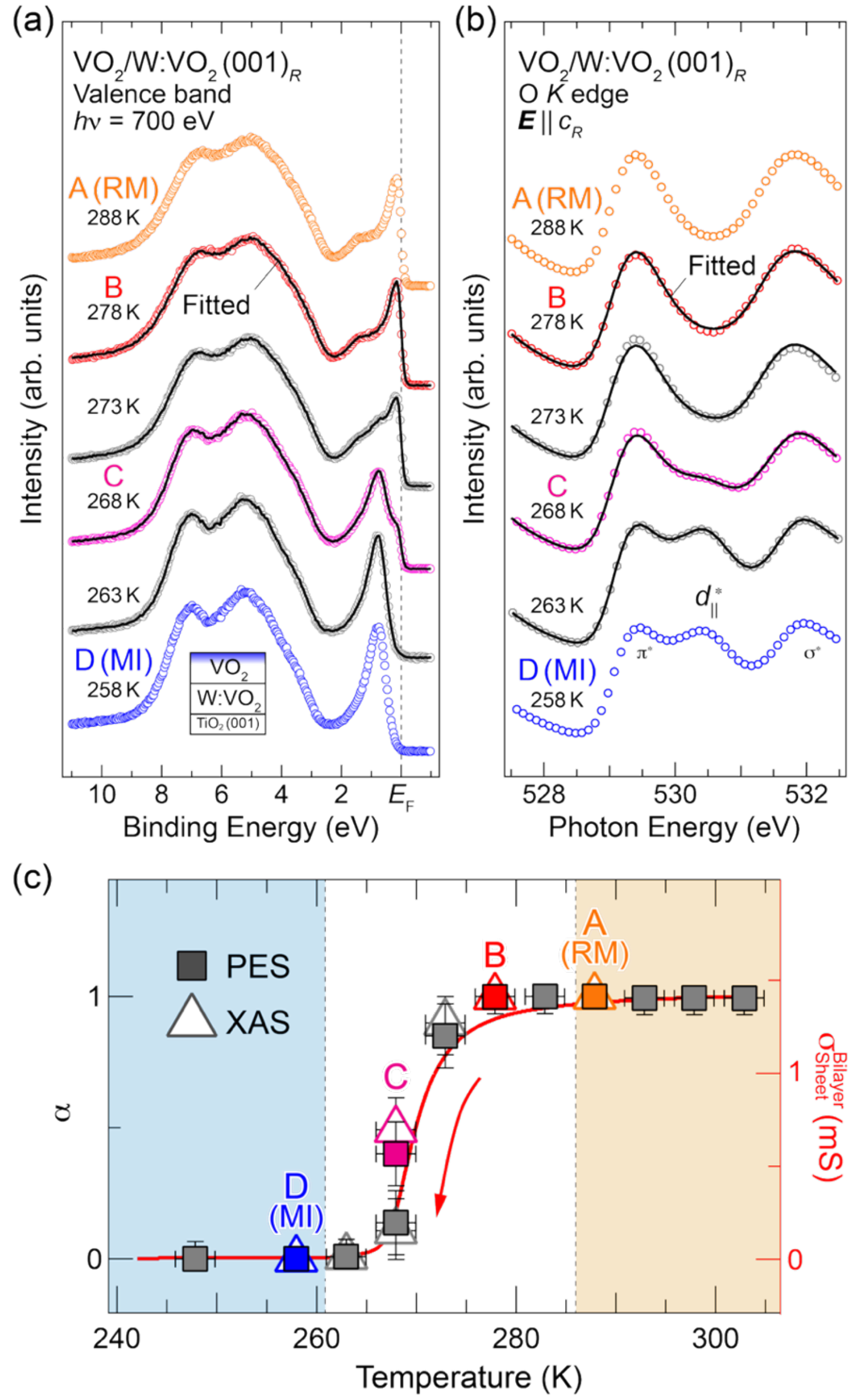

**FIG. 4.** Temperature dependence of *in situ* (a) valence-band and (b) O $K$ XAS spectra measured near $T_{\mathrm{MIT}}$ upon cooling for $VO_2$ (4.5 nm)/W:$VO_2$ (4.5 nm) bilayers. Note that the spectra mostly reflect the electronic structure of the upper $VO_2$ layer owing to the surface sensitivity of the present spectroscopic measurements. The fitted results by the linear combination of the rutile

metallic ($A$) and the monoclinic insulating ($D$) phases [Eq. (1)] are overlaid on respective spectra. The spectra are almost perfectly described by the linear combination. (c) Plot of $\alpha$ as a function of temperature. The $\sigma_{\mathrm{Sheet}}$ of the bilayer (Fig. 1) is overlaid as a red curve for comparison. Color shading corresponds to that in Figs. 1 and 2. MI and RM denote the monoclinic insulator and rutile metal, respectively.

The selective observation of the electronic and crystal structures of the constituent layers in the bilayers indicates that the upper $VO_2$ layer undergoes the transition from the monoclinic insulating to the rutile metallic phase by forming the heterointerface between $VO_2$ and $W{:}VO_2$ layers. The significance of this study is the experimental demonstration of a collective monoclinic insulating-to-rutile metallic phase transition induced in the $VO_2/W{:}VO_2$ heterostructures. By providing experimental validation for the proposed mechanism [25], the present study offers a solid basis for the thermodynamic modeling approach (including calculations of the free energy) and reveals the importance of such energy balance in bilayer systems. These results therefore establish a solid groundwork for future quantitative understanding of $VO_2$-based heterostructures. However, it should be bear in mind that the present study does not eliminate the possibility of the emergence of the monoclinic metallic phase in $VO_2$-based bilayer structures. In comparison to the previous results of Lee *et al*. [26], there are several differences from the bilayers examined here: the difference (7 K) of $T_{\mathrm{MIT}}$ between constituent layers is much smaller than that in the present case (25 K), and electrons are doped by oxygen vacancies ($VO_{2-\delta}$). The significantly smaller difference in $T_{\mathrm{MIT}}$ reflects a closer proximity between the electronic and structural energies of constituent layers, which may induce the emergence of the new equilibrium monoclinic metallic phase in the $VO_2/VO_{2-\delta}$ bilayer [26]. Furthermore, in chemically doped $VO_2$, the V-V dimer structure is known to change significantly depending on the type and concentration of the dopant, resulting in the complicated electronic phase diagram [32,50,51]. Therefore, the resultant complicated static energy balance between the interfacial energy and the bulk free energies of constituent layers may lead to the emergence of the monoclinic metallic phase under certain conditions. To gain a better understanding of the interface-induced collective phenomena occurring in the $VO_2$-based bilayers, further systematic investigations are required. In particular, investigations of the more detailed dopant and layer-thickness dependencies are necessary. Complementary probes such as temperature-dependent Raman spectroscopy may offer additional insights into the interplay between the electronic and structural transitions [52,53]. Also, further theoretical works that adequately treat these effects will be necessary to examine the possibility of a new electronic phase in the bilayers.

## V. CONCLUSION

We performed *in situ* PES and XAS measurements on $VO_2$/W:$VO_2$ bilayers to investigate the changes in the electronic structure and characteristic V-V dimerization across the collective phase transition induced at the heterointerface between the monoclinic insulating phase $VO_2$ and rutile metallic phase W:$VO_2$ layers. Thanks to the surface sensitivity of PES and XAS in the soft x-ray region, we separately extracted the changes in the top $VO_2$ layer. The spectra exhibited remarkable change associated with the collective phase transition: (1) The upper $VO_2$ layer exhibits almost the same spectral behavior across the MIT as that of a $VO_2$ single-layer film, whereas its $T_{MIT}$ is slightly lower than that of the single-layer film. (2) In the metallic states of the upper $VO_2$ layer, there is no indication of the V-V dimerization. (3) During the phase transition, both the PES and XAS spectra are described by a linear combination of the rutile metallic and monoclinic insulating phases, indicating the occurrence of in-plane phase separation. These results strongly suggest that the upper $VO_2$ layer undergoes a collective transition from the monoclinic insulating to the rutile metallic phase by forming the heterointerface with the electron-doped $VO_2$ layer. The occurrence of the phase transition from the monoclinic insulating to rutile metallic phase in the $VO_2$ upper layers suggests that the collective phase transition originates from the static energy balance between the interface energy and the bulk free energies of the constituent layers.

## DATA AVAILABILITY

The datasets generated and/or analyzed during the current study are available from the corresponding author upon reasonable request.

## ACKNOWLEDGMENTS

The authors are very grateful to T. Yajima, S. Biermann, and M. J. Rozenberg for our helpful discussions. This work was financially supported by a Grant-in-Aid for Scientific Research (Nos. 20KK0117, 21K20498, 22H01947, 22H01948, and 23H00263) from the Japan Society for the Promotion of Science (JSPS), CREST (JPMJCR18T1) from the Japan Science and Technology Agency (JST), and MEXT Program: Data Creation and Utilization Type Material Research and Development Project (Grant No. JPMXP1122683430). S.I. and T.K. acknowledge the financial support from the Division for Interdisciplinary Advanced Research and Education at Tohoku University. N.H. acknowledges the financial support from the Chemistry Personnel Cultivation Program of the Japan Chemical Industry Association. The work performed at KEK-PF was approved by the Program Advisory Committee (proposal Nos. 2019T004, 2022G675, 2021S2-002, and 2024S2-003) at the Institute of Materials Structure Science, KEK.

# Supplemental Material

## Interface-induced collective phase transition in $VO_2$-based bilayers studied by layer selective spectroscopy

D. Shiga[1,2,*], S. Inoue[1], T. Kanda[1], N. Hasegawa[1], M. Kitamura[2], K. Horiba[2], K. Yoshimatsu[1], A. F. Santander-Syro[3], and H. Kumigashira[1,2,*]

[1] *Institute of Multidisciplinary Research for Advanced Materials (IMRAM), Tohoku University, Sendai 980–8577, Japan*

[2] *Photon Factory, Institute of Materials Structure Science, High Energy Accelerator Research Organization (KEK), Tsukuba 305–0801, Japan*

[3] *Institut des Sciences Moléculaires d'Orsay, Université Paris-Saclay, 91405 Orsay, France*

[*]Correspondence authors: dshiga@tohoku.ac.jp, kumigashira@tohoku.ac.jp

## Supplemental Note 1. Sample characterization

### A. Surface crystallinity and morphology

The surface structures and cleanliness of all the measured samples — 9 nm $VO_2$ and $V_{0.99}W_{0.01}O_2$ (W:$VO_2$) single-layer films and $VO_2$ (4.5 nm)/W:$VO_2$ (4.5 nm) bilayer structures grown on $TiO_2$ (001) substrates — were confirmed via *in situ* reflection high-energy electron diffraction (RHEED). Typical *in situ* RHEED patterns are shown in the left panel of Fig. S1. The integer-order Bragg spots clearly appear at the same positions as those of the bare $TiO_2$ surface as indicated by vertical dashed lines. Furthermore, clear Kikuchi lines are also observed in the samples and substrates. These results indicate good surface crystallinity and cleanliness of the bilayers, as well as coherent epitaxial growth on the substrates.

The atomically flat surfaces of all the measured samples were confirmed by *ex situ* atomic force microscopy (AFM). The corresponding AFM images are shown in the right panel of Fig. S1. The AFM images show similar surface morphologies to that of the original $TiO_2$ substrate. The root-mean-square roughness $R_{rms}$ estimated from the AFM images is all less than 0.2 nm. These values are almost the same as that of the original $TiO_2$ substrate ($R_{rms}$ = 0.16 nm). The $R_{rms}$ values of the measured films were all less than the V-V dimer length (approximately 0.3 nm), indicating that these films were controlled to the scale of the V-V dimer length and that the smooth surface and interface were maintained not only in the single-layer films but also in the bilayer structures.

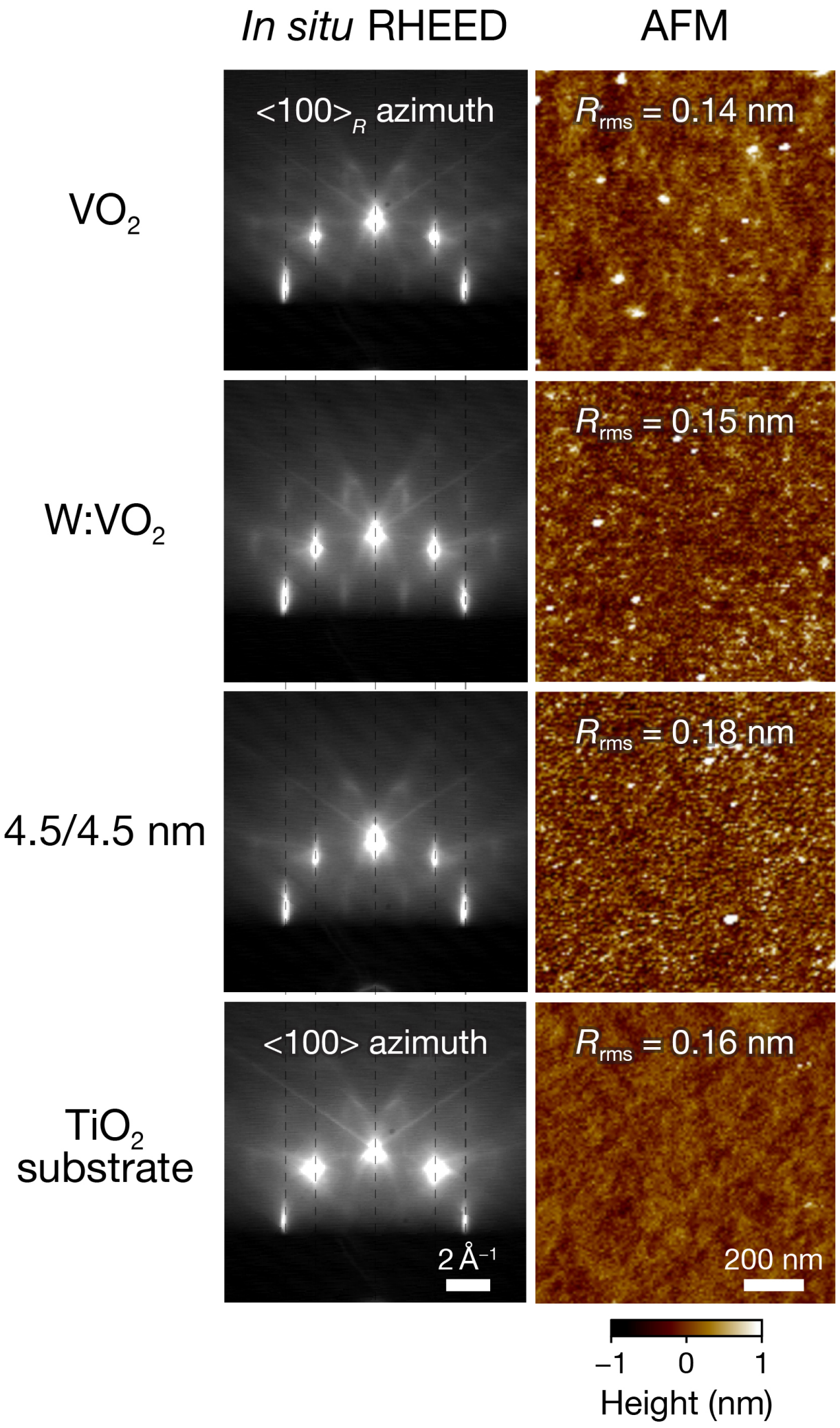

**Fig. S1.** Typical *in situ* RHEED patterns (0th Laue zone) captured along the <100>$_R$ azimuthal direction (left panel) and the corresponding AFM images (right panel) of the measured 9 nm $VO_2$ single-layer films, 9 nm W:$VO_2$ single-layer films, and $VO_2$ (4.5 nm)/W:$VO_2$ (4.5 nm) bilayer structures grown on $TiO_2$ (001) substrates, along with those of the substrate as references.

## B. Crystal structure

The crystal structures of the 9 nm $VO_2$ and $V_{0.99}W_{0.01}O_2$ (W:$VO_2$) single-layer films and the $VO_2$ (4.5 nm)/W:$VO_2$ (4.5 nm) bilayer structures were characterized by x-ray diffraction (XRD) measurements, which confirmed the achievement of single phase in both the single-layer films and the absence of another phase in the bilayer, as well as their coherent growth on $TiO_2$ (001) substrates. Figure S2(a) shows the out-of-plane XRD patterns around the $(002)_R$ reflection for the films and bilayer. The formation of single phase for the films and the absence of another phase in the bilayer are confirmed. In addition, the presence of well-defined Laue fringes indicates the formation of atomically flat surfaces and chemically abrupt interfaces. The estimated out-of-plane lattice constant ($c_R$) of the $VO_2$ and W:$VO_2$ films are 0.2829(4) nm and 0.2839(4) nm, respectively. These values are in good agreement with previous reports [32], guaranteeing the high quality of the present samples. The XRD pattern for the bilayer is well fitted with a two-layer model consisting of a $VO_2$ layer [$c_R$ = 0.2829(4) nm] and a W:$VO_2$ layer [$c_R$ = 0.2839(4) nm], indicating the formation of a chemically abrupt heterointerface between the two layers, as well as between the layer and the substrate, as reported previously [25]. The coherent growth of the bilayer on $TiO_2$ (001) substrates is confirmed by the reciprocal space mapping (RSM) around the $(112)_R$ reciprocal point, as well as the absence of another phase, as shown in Fig. S2(b). As shown in Fig. S2(b), the in-plane lattice constant of the layers remains pinned to that of the $TiO_2$ substrate, indicating the coherent growth of the bilayer. These crystallographic results identify the present films and bilayers as being highly crystalline.

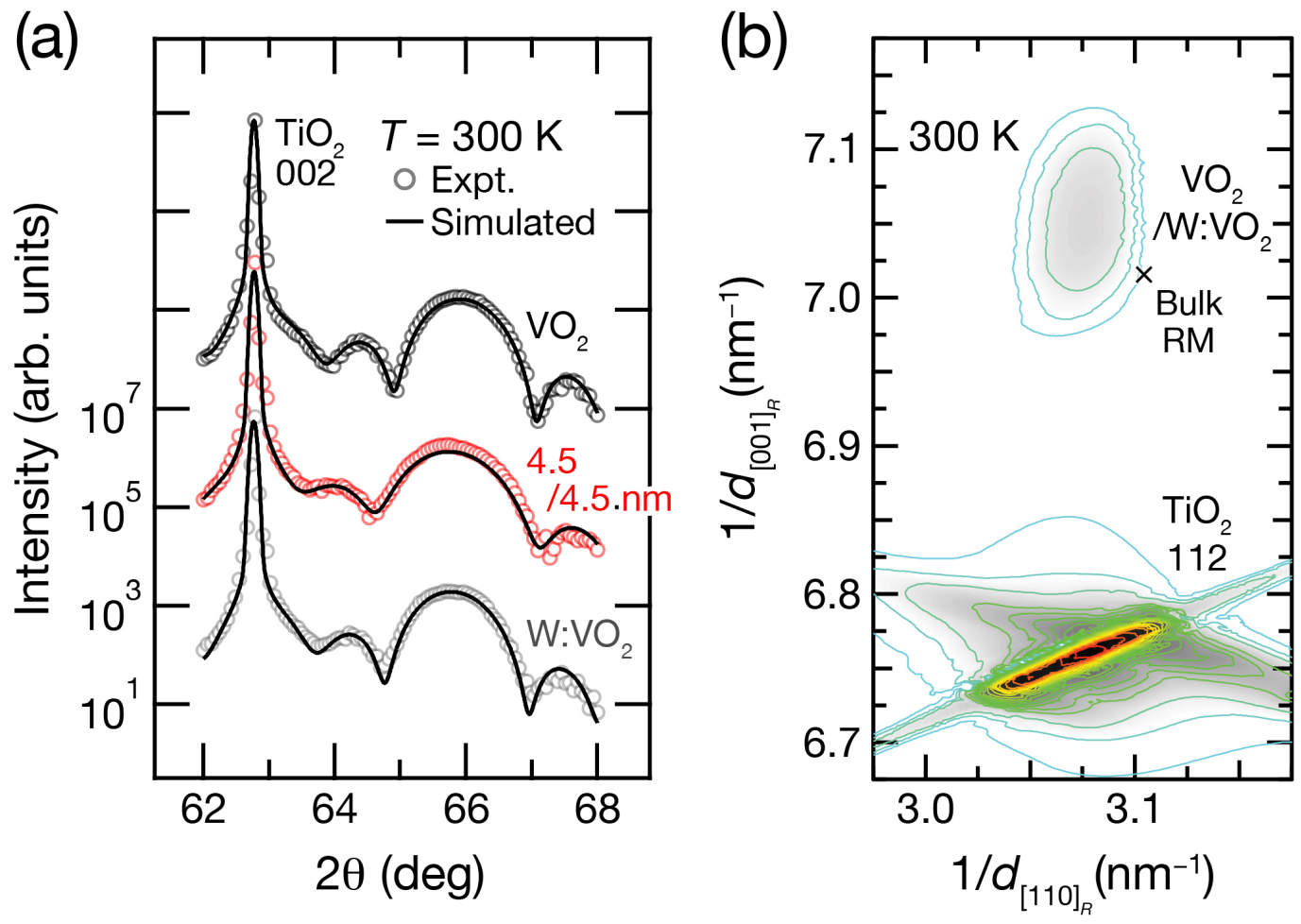

**Fig. S2.** (a) Typical out-of-plane XRD patterns around the $(002)_R$ reflection measured at room temperature for 9 nm $VO_2$ and W:$VO_2$ single-layer films and $VO_2$ (4.5 nm)/W:$VO_2$ (4.5 nm) bilayers grown on $TiO_2$ (001) substrates. Simulated XRD patterns based on corresponding single- or two-layer models (black curves) are overlaid. (b) RSM around the $(112)_R$ reciprocal point for the bilayer. The cross mark denotes lattice constants of bulk $VO_2$ in the rutile metallic (RM) phase [33] for reference.

## Supplemental Note 2. Core-level photoemission measurements

### A. Chemical states of V ions

Figure S3(a) shows the *in situ* V 2*p* core-level spectrum measured in the monoclinic insulating phase for the 4.5 nm bilayer structures, alongside hard x-ray photoemission (HAXPES) spectra for $VO_2$ single-layer films as a reference for the $V^{4+}$ state [15]. The V 2*p* core-level spectrum of the bilayer is almost identical to that of $V^{4+}$. The excellent agreement indicates that the V ions in the bilayers exist in a tetravalent state.

### B. Formation of a chemically abrupt interface

Figure S3(b) shows the *in situ* V 3*p* and W 4*f* core-level spectra for the 6.5 and 4.5 nm bilayers, as well as the 9 nm W:$VO_2$ single-layer films. Although the doping concentration is only 1at% ($x = 0.01$ in $V_{1-x}W_xO_2$), prominent W 4*f* peaks are clearly observed in the 9 nm W:$VO_2$ single-layer films due to the sharpness and large photoionization cross section of the W 4*f* states. This enables us to evaluate the degree of interdiffusion of W ions in the bilayer structures using the attenuation of the W 4*f* signal. As can be seen in Fig. S3(b), with increasing the undoped $VO_2$ overlayer thickness ($t$), the intensity of W 4*f* states ($I_{W\,4f}$) steeply reduces and almost disappears at $t = 6.5$ nm, owing to the attenuation of photoelectrons emitted from the buried W:$VO_2$ layer by the $VO_2$ overlayer. This steep attenuation behavior strongly suggests the formation of a chemically abrupt interface. To evaluate the chemical abruptness of the interfaces, we plot $I_{W\,4f}$ as a function of $t$ in the inset of Fig. S3(b) and compare it with a photoemission attenuation function calculated using the TPP-2M code [34]. The excellent agreement between the two indicates the formation of a chemically abrupt interface (i.e., negligible interdiffusion of W ions) in the present bilayers.

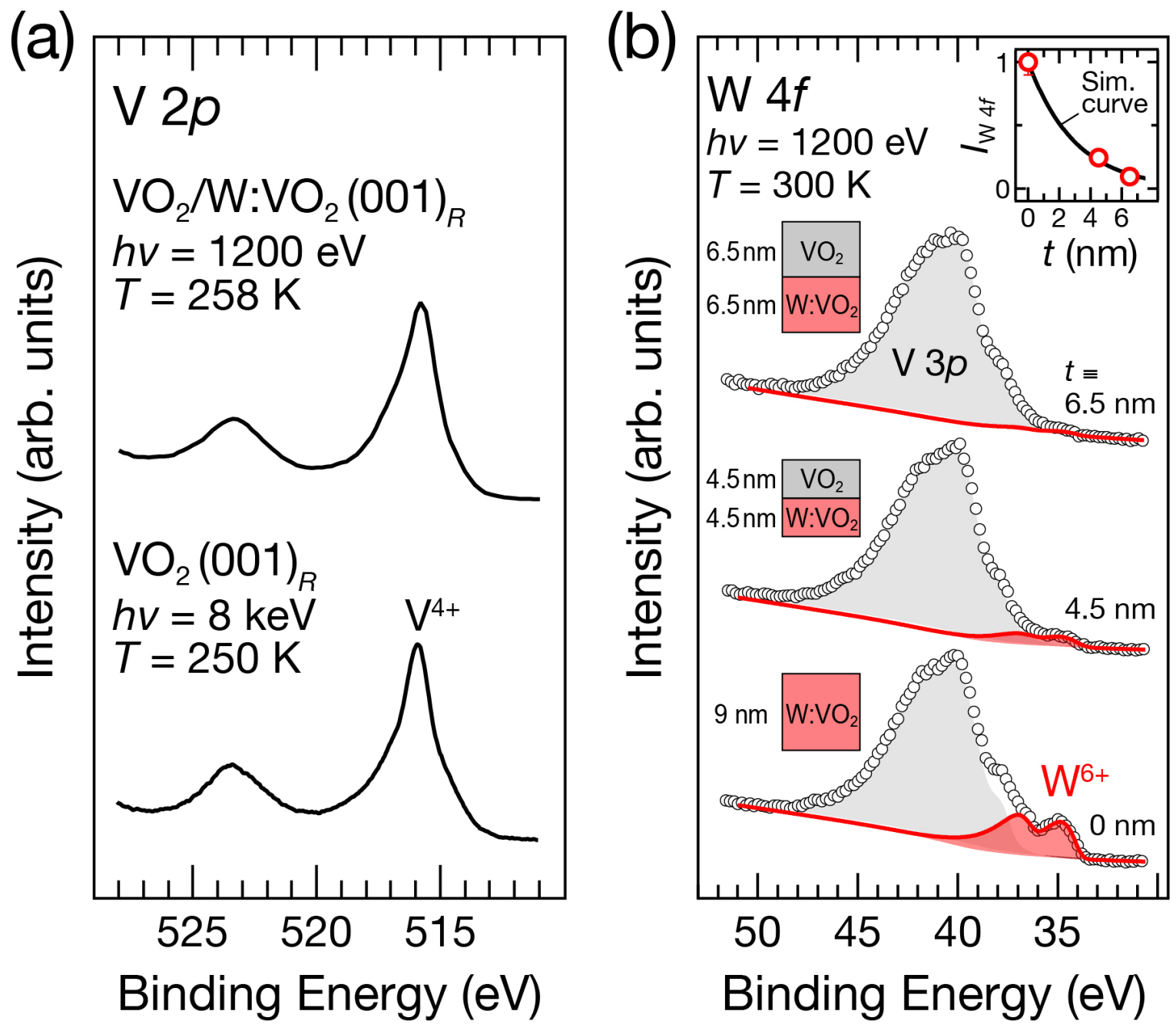

**Fig. S3.** (a) *In situ* V 2*p* core-level spectrum measured at $T$ = 258 K (insulating phase) for the $VO_2$ (4.5 nm)/W:$VO_2$ (4.5 nm) bilayer structures, alongside HAXPES spectrum for $VO_2$ single-layer films as a reference for the $V^{4+}$ state [15]. (b) V 3*p* and W 4*f* core-level spectra for the 6.5 nm and 4.5 nm bilayers and 9 nm W:$VO_2$ single-layer films. Each spectrum is fitted with Voigt functions corresponding to the relevant core levels after subtracting the background using a Shirley function. Inset shows a plot of $I_{W\,4f}$ (red open circles) as a function of *t*, in comparison with a calculated photoemission attenuation curve (black line) based on the TPP-2M code [34]. The excellent agreement between the two indicates the formation of a chemically abrupt interface in the present bilayers.

## Supplemental Note 3. Experimental geometry in polarization-dependent x-ray absorption measurement

Figure S4 depicts the schematic of our experimental geometry for *in situ* polarization-dependent x-ray absorption spectroscopy (XAS) measurements, including the crystal axes of a $VO_2$ $(001)_R$ film sample and the polarization vector $\boldsymbol{E}$. Regarding linear dichroism measurements for XAS, we acquired the spectra at angles $\theta = 0°$ and 60° between $\boldsymbol{E}$ and the $a_R$-axis direction, which is defined as the $a$-axis direction in the rutile structure ($[100]_R$), while maintaining a fixed angle of 60° between the direction normal to the $(001)_R$ surface and the incident light. The maintenance of the fixed angle between the direction normal to the surface and the incident light ensures that the probing depth corresponding to the two spectra with different $\theta$ values is the same. In the present experimental geometry, XAS spectra with $\boldsymbol{E} \parallel c_R$ ($I_\parallel$) can be deduced from the expression $I_\parallel = (4/3)(I - I_\perp/4)$, where $I$ and $I_\perp$ (namely, that corresponding to $\boldsymbol{E} \perp c_R$) denote XAS spectra measured with grazing ($\theta = 60°$) and normal ($\theta = 0°$) incidences, respectively.

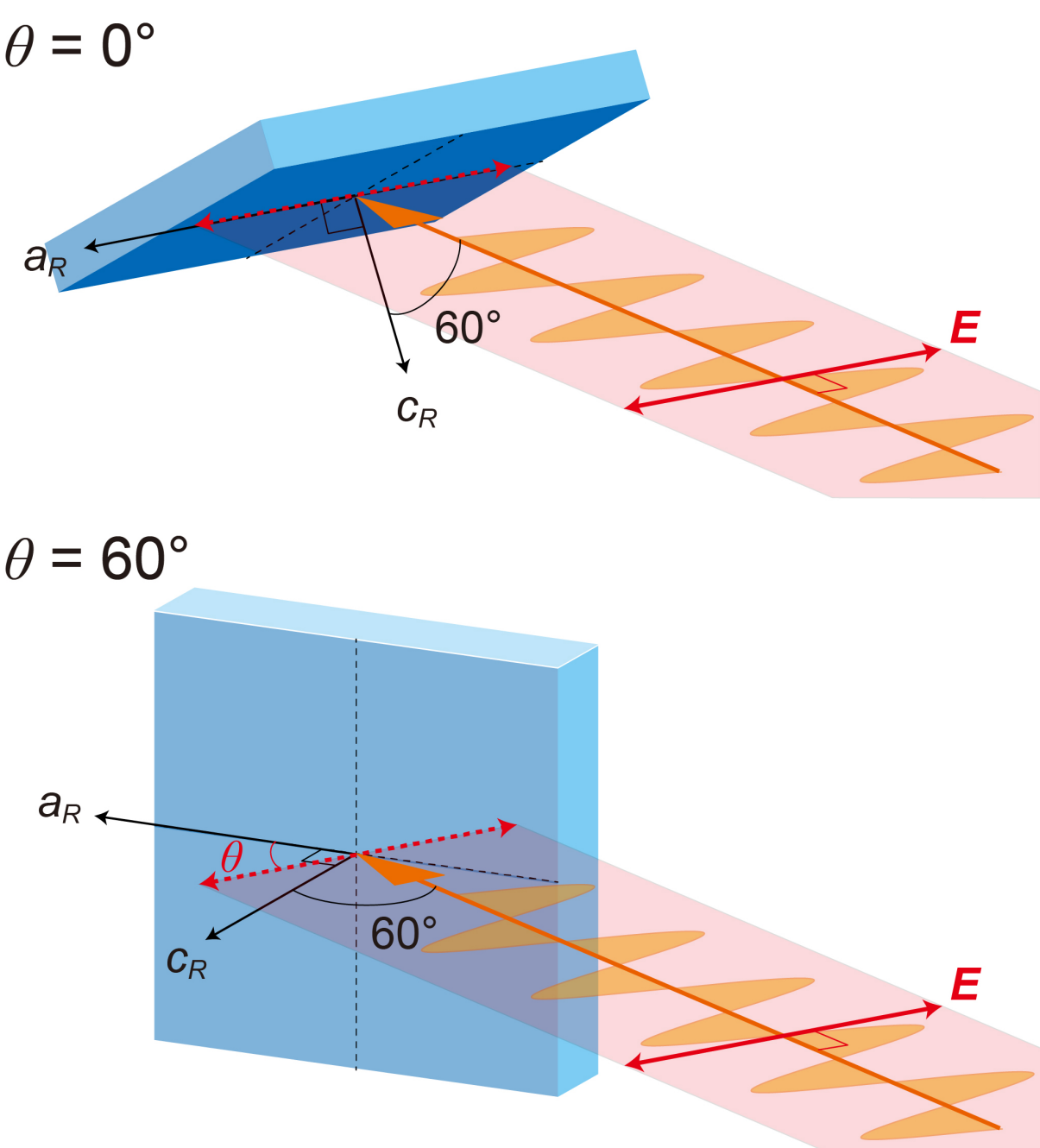

**Fig. S4.** Schematic of our experimental geometry of polarization-dependent XAS measurements of a $VO_2$ $(001)_R$ film at $\theta = 0°$ and 60°. The $a_R$- and $c_R$-axis directions are defined as the $a$- and $c$-axis directions in the rutile structure, respectively.

## Supplemental Note 4. Probing depths of the present soft x-ray spectroscopies

To evaluate the probing depth of photoemission spectroscopy (PES) $\lambda^{\mathrm{PES}}$ and x-ray absorption spectroscopy (XAS) $\lambda^{\mathrm{XAS}}$, we measured the overlayer-thickness dependence of Ti 2$p$ core-level spectra and Ti $L_3$-edge XAS for $VO_2$/Nb:$TiO_2$ (001) single-layer films, as shown in Figs. S5(a) and S4(b), respectively. Each spectrum is normalized to the incident photon flux; hence, the observed intensity reduction with increasing $VO_2$ overlayer thickness $t$ reflects the attenuation of the Ti-derived signal from buried $TiO_2$ substrates by the $VO_2$ overlayer. For both the soft x-ray (SX) spectroscopies, the intensity decreases steeply with increasing $t$ and becomes almost undetectable at $t = 3$ nm, highlighting the surface sensitivity of the two spectroscopic measurements, as well as the formation of the chemically abrupt interface between the $VO_2$ layer and $TiO_2$ substrate [15]. To evaluate the probing depths for PES and XAS, we plot the Ti 2$p$ core-level intensity $I_{\mathrm{Ti}\,2p}^{\mathrm{PES}}$ and Ti $L_3$ XAS intensity $I_{\mathrm{Ti}\,2p}^{\mathrm{XAS}}$ as a function of $t$ in Fig. S5(c) and fit them by attenuation functions of $I_{\mathrm{Ti}\,2p}^{\mathrm{PES}} = e^{-t/\lambda^{\mathrm{PES}}}$ for PES and $I_{\mathrm{Ti}\,2p}^{\mathrm{XAS}} = e^{-t/\lambda^{\mathrm{XAS}}}$ for XAS, respectively. As shown in Fig. S5(c), the intensity reductions are well fitted to the attenuation functions with $\lambda^{\mathrm{PES}} = 0.55(2)$ and $\lambda^{\mathrm{XAS}} = 0.70(1)$. The similar values of $\lambda^{\mathrm{PES}}$ and $\lambda^{\mathrm{XAS}}$ indicate that both SX spectroscopies probe approximately the same region. Based on these values, the signal contributions from the upper 4.5 nm thick $VO_2$ layer in the present PES and XAS measurements are estimated to be 99.97% and 99.84%, respectively. These results indicate that both PES and XAS measurements predominantly probe the upper $VO_2$ layer of bilayer structures.

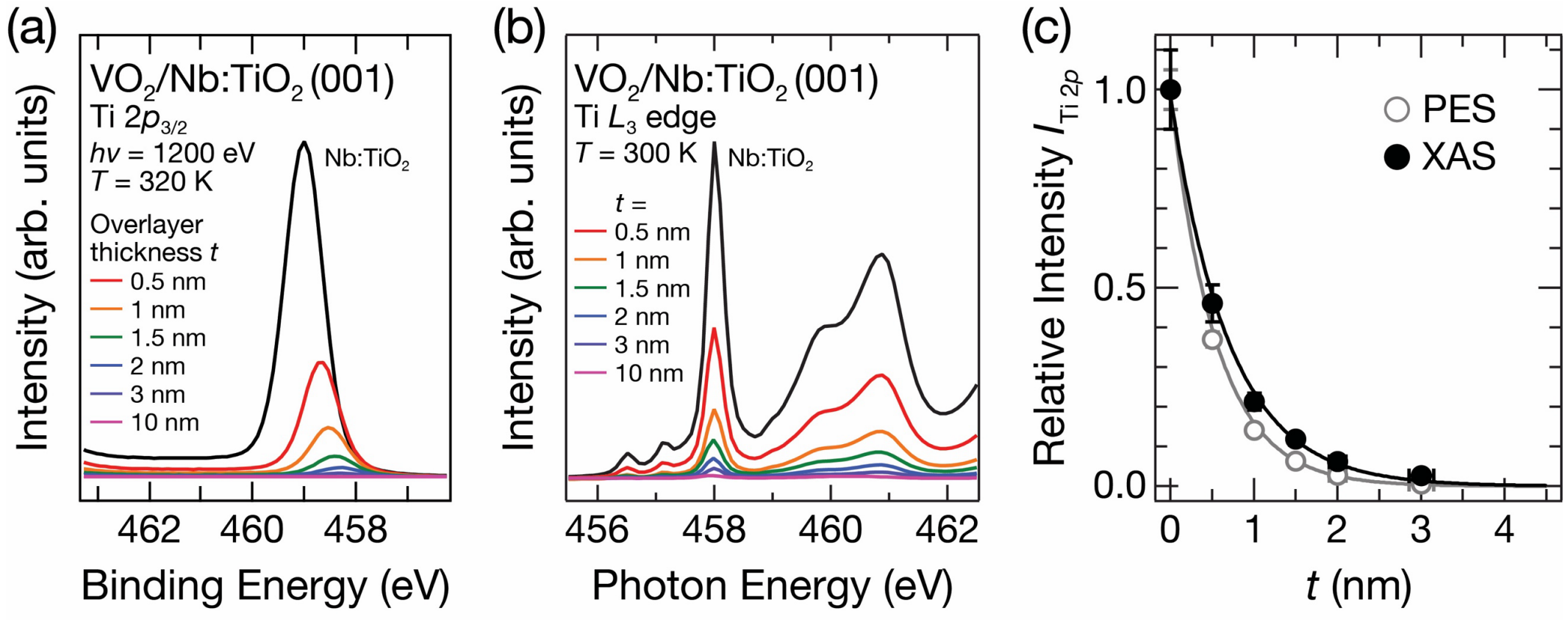

**Fig. S5.** (a) Ti 2$p_{3/2}$ core-level spectra [15] and (b) Ti $L_3$ XAS spectra for $VO_2$/Nb:$TiO_2$ (001) single-layer films with varying $t$, along with those of the Nb:$TiO_2$ substrate as references. (c) Relative intensities of the Ti 2$p_{3/2}$ core-level PES spectra ($I_{\mathrm{Ti}\,2p}^{\mathrm{PES}}$, gray open circles) and Ti $L_3$

XAS spectra ($I^{\mathrm{XAS}}_{\mathrm{Ti}\ 2p}$, black solid circles), plotted as functions of $t$. Gray and black curves represent fits using attenuation models with $\lambda^{\mathrm{PES}}$ = 0.55(2) nm and $\lambda^{\mathrm{XAS}}$ = 0.70(1) nm, respectively.

## Supplemental Note 5. Irradiation-time dependence of valence band spectra

It is well known that $VO_2$ exhibits an insulator-to-metal transition upon irradiation by soft x-ray (SX) [35]. Therefore, to perform the photoelectron spectroscopic measurements within a period in which no detectable spectral changes due to SX irradiation are observed, we measured the irradiation-time dependence of the valence-band spectra for insulating $VO_2$, as shown in Fig. S6. No detectable spectral differences near the Fermi level ($E_F$) are observed between the spectra acquired before (initial) and after 15 minutes of SX irradiation, whereas certain slight changes are detectable in the spectra after 45 minutes. Therefore, we changed the location of the light spot on each sample every 15 minutes during the present measurements. All the PES and XAS spectra shown in the text and Supplemental Material were acquired within this period. Thus, the light irradiation effects are negligible in the present study.

As shown in Fig. S6, the intensity (area under the curve) of the O 2*p* bands decreases after 45 minutes of SX irradiation in parallel with the broadening of the V 3*d* band near $E_F$, whereas the area under the curve for the V 3*d* band remains unchanged. This indicates that the changes in the V 3*d* spectra near $E_F$ might originate from oxygen vacancies generated by SX irradiation, as observed in many transition metal oxide systems [35–38].

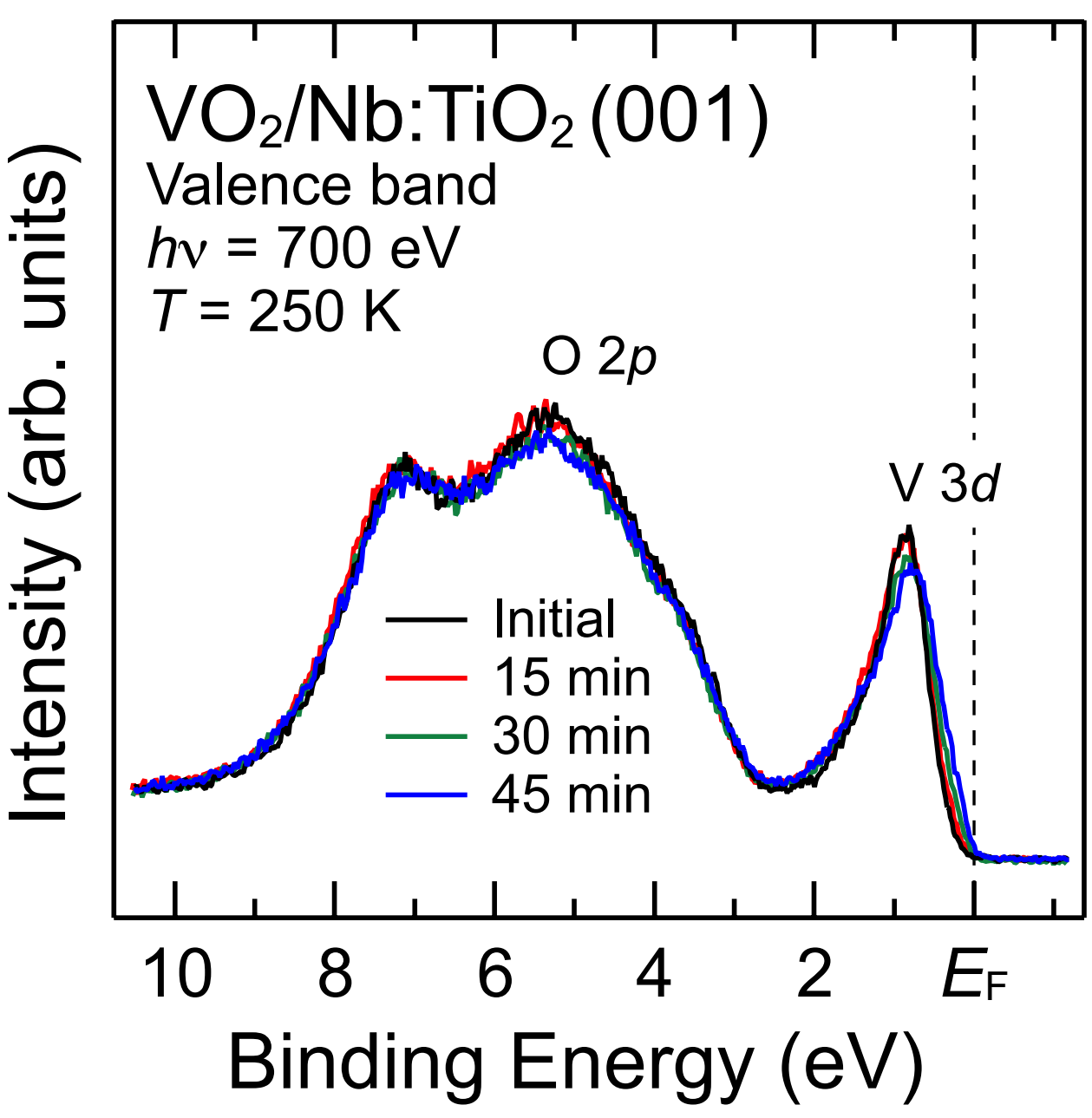

**Fig. S6.** Irradiation-time dependence of valence-band spectra for $VO_2$/Nb:$TiO_2$ (001) films.

## Supplemental Note 6. Thermal hysteresis in sheet resistance

Figure S7 shows the temperature dependence of the sheet resistance $R_{Sheet}$ for $VO_2$ (4.5 nm)/W:$VO_2$ (4.5 nm) bilayers, along with 9 nm epitaxial $VO_2$ and W:$VO_2$ single-layer films. Note that the $R_{Sheet}$-$T$ curves measured upon cooling are the same as those in Fig. 2 in the main text. The hysteresis loop characteristic of a first-order transition in $VO_2$ films is clearly observed for the bilayer. The hysteresis loops almost close at 300 K. Therefore, to avoid the possible hysteresis effects in the present spectroscopic measurements, the sample temperature was maintained at 320 K for half an hour prior to the spectroscopic measurements, and then all spectroscopic data were acquired only during cooling.

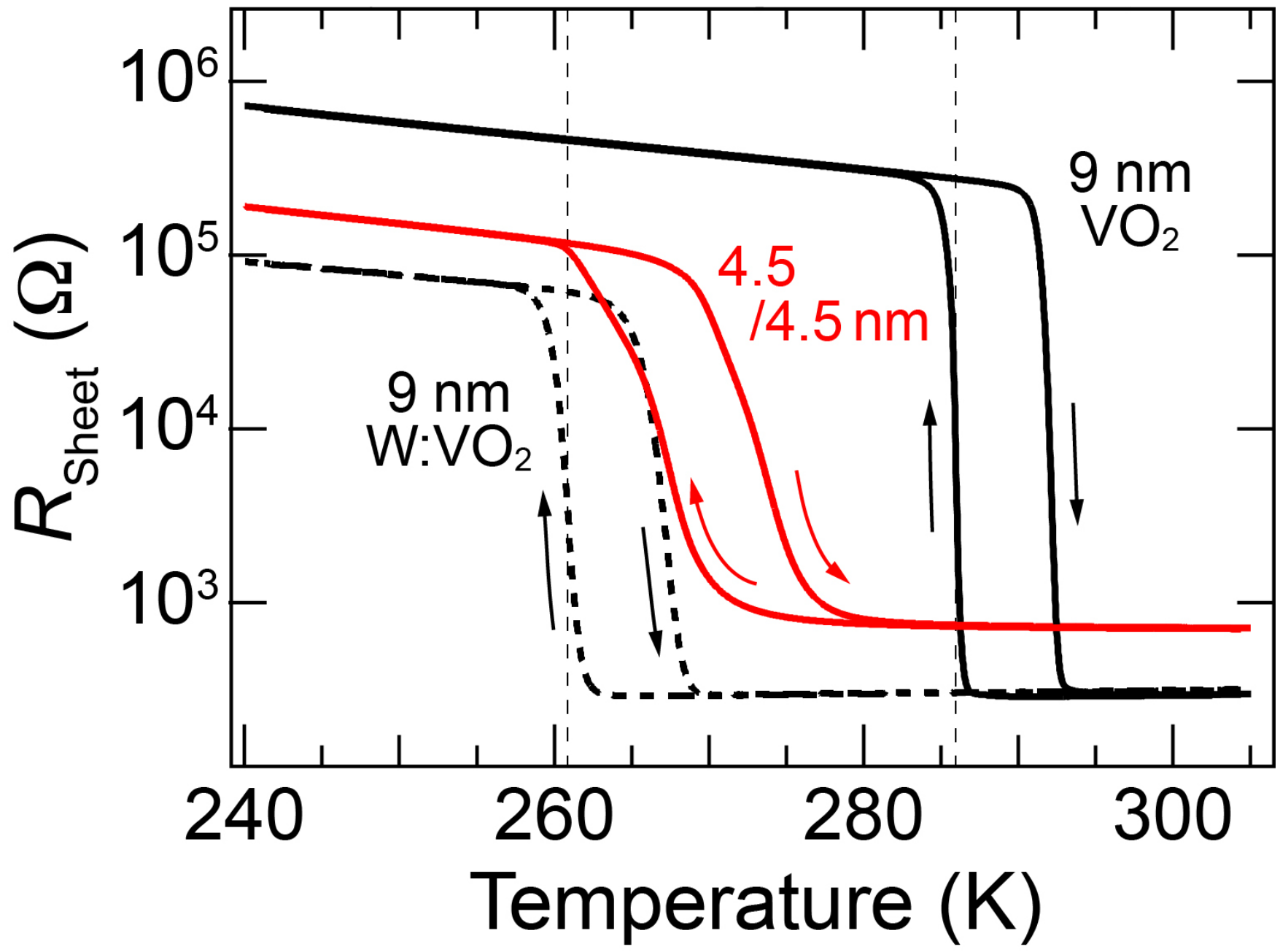

**Fig. S7.** Temperature dependence of $R_{Sheet}$ for 4.5/4.5 nm bilayer, which consists of $VO_2$ (4.5 nm) on W:$VO_2$ (4.5 nm) (solid red curve), and 9 nm epitaxial $VO_2$ and W:$VO_2$ single-layer films (solid and dashed black curves, respectively) grown on $TiO_2$ (001) substrates. Vertical dashed lines indicate $T_{MIT}^{Cooling}$ of the $VO_2$ (286 K) and electron-doped W:$VO_2$ (261 K) single-layer films.

## Supplemental Note 7. Definition of the interface-induced transition

In the present bilayer structures, we observed that the upper $VO_2$ layer, which is originally in the monoclinic insulating phase, undergoes a transition to the rutile metallic phase upon forming the interface with the W:$VO_2$ layer at temperature *B*, as schematically illustrated in the inset of Fig. 2. In this context, the phase transition is not triggered by temperature but rather driven by interface effects. In other words, temperature merely serves as an external tuning parameter to observe the changes induced by interface effects, but it is not the critical driving factor for the interface-induced phase transition. Therefore, although the observed phenomena can indeed be considered an integrated effect of both temperature- and interface-driven transitions, we define it as an "interface-induced transition" in a narrow sense, to distinguish it from the usual temperature-driven transition.